\documentclass[namedreferences]{SolarPhysics}
\usepackage[optionalrh]{spr-sola-addons} 
\usepackage{graphicx}                    
\usepackage{color}                       
\usepackage{url}                         

\graphicspath{%
    {converted_graphics/}
    {C:/aa/ddr/stereo/HET/}
}
\begin{document}

\begin{article}

\begin{opening}

\title{North/South Hemispheric Periodicities in the $>25$~MeV Solar Proton Event Rate During the Rising and Peak Phases of Solar Cycle 24
}
%
\author{I.~G.~\surname{Richardson}$^{1, 2}$\sep
        T.~T.~\surname{von Rosenvinge}$^{2}$\sep
        H.~V.~\surname{Cane}$^3$
       }


%
\runningauthor{I. G. Richardson et al.}
\runningtitle{$\approx6$ Six--Month SEP Periodicity}

%
  \institute{$^{1}$CRESST and Department of Astronomy, University of Maryland, College Park, Maryland, 20742, USA\\
email: \url{ian.g.richardson@nasa.gov}\\
$^{2}$Code 661, NASA Goddard Space Flight Center, Greenbelt, Maryland, 20771, USA\\
                     email: \url{tycho.t.vonrosenvinge@nasa.gov}\\
             $^3$Department of Mathematics and Physics, University of Tasmania, Hobart, Tasmania, Australia\\
email: \url{hilary.cane@utas.edu.au}\\
$\copyright$ 2016. All rights reserved. 
             }

\begin{abstract}
We present evidence that $>25$~MeV solar proton events show a clustering in time at intervals of $\approx$ six months that persisted during the rising and peak phases of Solar Cycle 24. This phenomenon is most clearly demonstrated by considering events originating in the northern or southern solar hemispheres separately.  We examine how these variations in the solar energetic particle (SEP) event rate are related to other phenomena, such as hemispheric sunspot numbers and areas, rates of coronal mass ejections, and the mean solar magnetic field. Most obviously, the SEP event rate closely follows the sunspot number and area in the same hemisphere. The $\approx$ six-month variations are associated with features in many of the other parameters we examine, indicating that they are just one signature of the episodic development of Cycle 24. They may be related to the ``$\approx$150~day" periodicities reported in various solar and interplanetary phenomena during previous solar cycles. The clear presence of $\approx$six-month periodicities in Cycle 24 that evolve independently in each hemisphere conflicts with a scenario suggested by McIntosh {\it et al.} (2015, {\it Nature Com.} {\bf 6}, 6491) for the variational time scales of solar magnetism.

\end{abstract}

%
\end{opening}

%

\section{Introduction}

The starting point, and motivation for this study, is the catalog of solar energetic particle (SEP) events including 25~MeV protons compiled by \inlinecite{r14} using observations from the {\it Solar Terrestrial Relations Observatory Ahead and Behind} spacecraft (STEREO A and B: \opencite{k08}) and near-Earth spacecraft, since launch of the STEREO spacecraft on 26~October 2006.  \inlinecite{r14} noted that, during the rise phase of Solar Cycle 24 (which commenced in December 2008 based on the smoothed sunspot number), clusters of SEP events occurred at intervals of $\approx6$--7 months, separated by periods with few SEPs.  They suggested that this phenomenon might be related to the periodicities in SEP occurrence and other solar parameters reported in previous solar cycles ({\it e.g.}, \opencite{r84}; \opencite{crr98}; \opencite{d01}; \opencite{rc05}; and references therein). 

\begin{figure}
  \centering
 \includegraphics*[width=1.0\textwidth,angle=0]{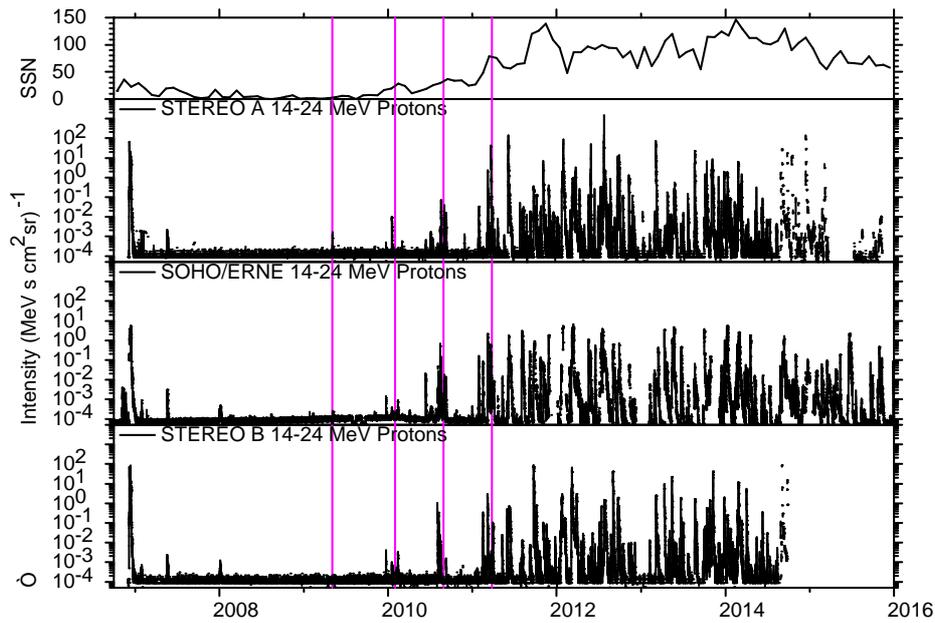}
  \caption{14\,--\,24~MeV proton intensities observed at STEREO A, B (second and bottom panels), and near Earth (third panel), from STEREO launch to the end of 2015.  The top panel shows the revised sunspot number from WDC-SILSO.  Vertical purple lines indicate the clustering of SEP events in time during the early rise phase of Solar Cycle 24.  The separation is nine months between the first two lines, and seven months between the other lines.}
  \label{3scsum0615}
\end{figure}

Figure~\ref{3scsum0615} is an updated version of Figure~1 of \inlinecite{r14} which illustrates this point.  The figure shows 14--24~MeV proton intensities observed by the {\it High Energy Telescopes} \cite{vr08} on STEREO A (second panel) and B (bottom panel) and by the {\it Energetic and Relativistic Nuclei and Electron} instrument (ERNE: \opencite{t95}) on the {\it Solar and Heliospheric Observatory} (SOHO) near Earth (third panel).  The top panel shows the recently-revised monthly-averaged sunspot number (from the World Data Center for Sunspot Index and Long-term Solar Observations (WDC-SILSO) at the Royal Observatory of Belgium, Brussels) which indicates the development of Solar Cycle 24.  The interval shown extends from STEREO launch to the end of 2015.  The STEREO A and B spacecraft moved ahead of or behind Earth in its orbit, respectively, advancing at $\approx$22$^{\circ}$ {\it per} year, and were above the west and east limbs of the Sun in February 2011.  By the end of the time interval shown, they had recently passed each other on the far side of the Sun; STEREO A was 169$^{\circ}$ east of the Earth and STEREO B was 175$^{\circ}$ to the west. Contact was unexpectedly lost with STEREO B on 1~October 2014, so no data beyond this date are shown in Figure~\ref{3scsum0615}.  STEREO A data are intermittent from mid~2014 due to heating issues when the spacecraft antenna was directed close to the Sun.  The spacecraft was also in safe mode in mid~2015 when passing directly behind the Sun as viewed from Earth.  

As discussed by \inlinecite{r14}, by examining SEP observations from the STEREOs and spacecraft (SOHO, the {\it Advanced Composition Explorer} (ACE), {\it Wind}, and the {\it Geostationary Operational Environmental Satellites} (GOES)) near Earth, and using solar imaging from the STEREOs and near-Earth spacecraft, individual SEP events observed at one or more locations, and their solar sources, may be identified unambiguously, even when the events originated on the far side of the Sun relative to Earth. In Figure~\ref{3scsum0615}, the large proton events in December 2006, just after STEREO launch and late in Cycle~23 ({\it e.g.}, \opencite{m08}; \opencite{m09}; \opencite{vr09}) were followed by an extended period with few particle enhancements during the solar minimum between Cycles 23 and 24.   The first 25~MeV proton event of Cycle 24 detected at both STEREOs and at Earth occurred on 22~December 2009 when STEREO A and B were $\approx65^o$ in longitude ahead of and behind the Earth, respectively  ({\it cf.}, Figure~13 of \opencite{r14}). This event occurred in the second of possibly as many as four brief intervals of enhanced SEP occurrence, indicated by purple vertical lines in Figure~\ref{3scsum0615}, separated by periods with relatively few SEP events, observed during the rise phase of Cycle~24. The first two purple lines are drawn separated by nine months, while the remainder are at seven month intervals, suggesting (as noted by \opencite{r14}) that the intervals of enhanced SEP occurrence are a quasi-periodic feature. It is unclear, however, from the particle intensity-time profiles in Figure ~\ref{3scsum0615} whether the SEP events continued to exhibit any periodic behavior further into the cycle as the SEP rate increased.  The main aim of this paper is to investigate whether this is the case.  

While discussing Figure~\ref{3scsum0615}, another feature to note is the interval from late 2012 until early 2013, between the two sunspot peaks in this cycle (top panel), that is characterized by a temporary decrease in the occurrence of large SEP events.  Although the ERNE data do have a data gap during part of this period, this feature is also evident at both STEREO spacecraft and in other near-Earth SEP data not shown here.  It is also apparent in Figure~\ref{3scsum0615} that the vast majority of the SEP events occurred in and after 2011, when the STEREO spacecraft were located above the far side of the Sun as viewed from Earth.  Hence, the STEREOs and spacecraft at Earth together provided an extended view in longitude at $\approx1$~AU of these SEP events ({\it e.g.}, \opencite{rou11}; \opencite{d12}; \opencite{lar13}; \opencite{w13}; \opencite{coh14}; \opencite{r14}, \opencite{pap14}; \opencite{lar14}; \opencite{gom15}; \opencite{vr15}; \opencite{lar16}).    

Although Figure~\ref{3scsum0615} shows 14\,--\,24~MeV proton intensities, the SEP event identifications in this article are based, as in \inlinecite{r14}, on the presence of 25~MeV protons. Table~1 of \inlinecite{r14} lists 209 individual events that included 25~MeV protons (detectable above instrumental backgrounds of $\approx10^{-4}$ (MeV~s~cm$^2$~sr)$^{-1}$) and their solar sources, identified from STEREO launch to the end of 2013 that are included in the present study.  Around 30\% of these events were observed by one or more STEREO spacecraft but not at the Earth.  Hence, the SEP event detection rate at $\approx1$~AU is increased by a factor of around a half over that at the Earth by including the STEREO observations.  A further 64 similar events in 2014 observed at the STEREOs and/or at Earth are also included in this study.

\section{Hemispheric SEP Rates}

\begin{figure} 
  \centering
 \includegraphics*[width=1.0\textwidth,angle=0]{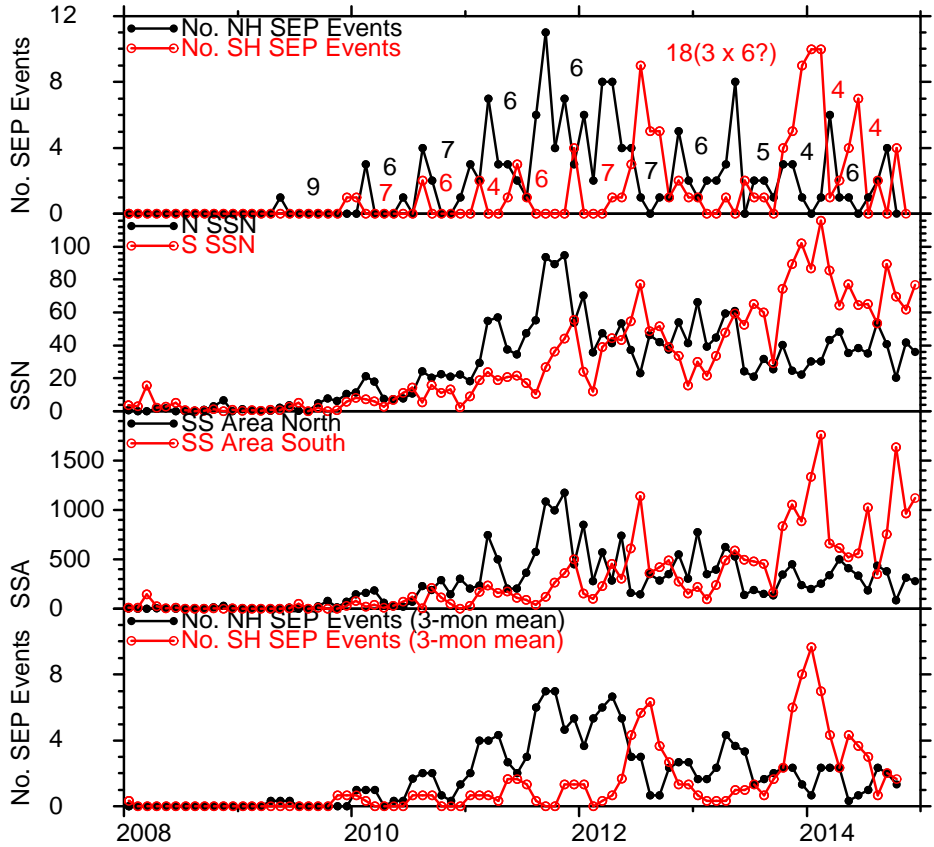}
  \caption{Comparison of the number of 25~MeV proton events {\it per} month in 2008\,--\,2014 detected at the STEREO spacecraft and/or at Earth (top panel) that originate in the northern (black) and southern (red) solar hemispheres with the monthly hemispheric sunspot numbers (SSN, second panel) and Greenwich - USAF/NOAA sunspot areas (SSA, third panel in millionths of the visible solar hemisphere  or MH).  Numbers in the top panel indicate the separation in data points ({\it i.e.}, months) between prominent peaks in the proton event occurrence rate in the respective hemispheres.  This is generally six or seven months in each hemisphere, but the variations in rate are independent in each hemisphere and closely follow those in the respective hemispheric sunspot number and area.  The bottom panel shows three-month running means of the hemispheric SEP occurrence rates.}
  \label{hemi}
\end{figure}

To help assess whether the SEP event clustering in Figure~\ref{3scsum0615} during the early rise phase of Cycle~24 persisted further into the cycle, the top panel of Figure~\ref{hemi} shows the SEP event occurrence (or more precisely, detection) rate, defined as the number of 25~MeV proton events detected/month, based on the 273 individual 25~MeV proton events identified from STEREO and near-Earth observations up to the end of 2014 discussed above.  In particular, we show separate rates for SEP events that originated in the northern (solid black-circles) or southern (open red-circles) solar hemispheres.  The source hemisphere is based on the helio-latitude of the associated flare, if present, for front side events, and from examining SOHO, {\it Solar Dynamics Observatory}, and STEREO movies of the solar activity associated with the SEP event in order to determine whether this activity occurred in the northern or southern active region belt.  For a handful of events, both northern and southern active regions were involved and the assignment was then based on the region where activity appeared to be initiated.  For comparison with the hemispheric SEP rates, the second panel of Figure~\ref{hemi} shows the northern (black) and southern (red) monthly-averaged revised sunspot numbers.  Note that the first sunspot maximum in Cycle 24, in 2011, was associated with northern hemisphere sunspots, while the second maximum, in late 2013\,--\,early 2014, was associated with southern sunspots.

Focusing first on the northern hemisphere SEP event rate (black graph in the top panel of Figure~\ref{hemi}), this is characterized by brief intervals of enhanced SEP occurrence separated by periods with lower rates.  Furthermore, this pattern is remarkably regular. The black numbers indicate the number of months ({\it i.e.}, data points) between the most prominent peaks in the SEP rate.  Although in a few cases, the peaks extend over more than one data point and some small secondary peaks are overlooked, the intervals between peaks are either six or seven months ($\approx183$ or 213~days) from the beginning of 2010, to late 2013.  After this time, intervals of 4\,--\,6 months are found.  

Considering the southern SEP event rate (red graph in the top panel of Figure~\ref{hemi}), again there are brief intervals of enhanced SEP occurrence separated by periods with few SEPs.  The separations between peaks (red numbers) are likewise generally 6\,--\,7 months.  It is, however, also evident that the SEP rates vary independently in both phase and amplitude in each hemisphere.  Early in the cycle, the rates in both hemispheres were nearly in phase.  However, following a four-month interval between southern hemisphere peaks in early 2011, the variations in each hemisphere moved to be in near anti-phase, a situation that persisted to around the middle of 2014.  Also, northern hemisphere SEPs were dominant from the beginning of the cycle to around mid~2013 except for an $\approx$ three-month interval in mid~2012 when southern events were temporarily dominant. Southern events were dominant again in late 2013\,--\,early 2014.  Interestingly, there were only minor southern hemisphere SEP rate peaks between these two prominent southern peaks, an interval of 18 months, or $3\times6$~months.  This suggests that southern SEP rate peaks were weaker, or maybe even absent, for two cycles of the $\approx$~six-month periodicity.  This feature is also associated with the interval of fewer SEPs in late 2012\,--\,early 2013 between the northern and southern sunspot peaks noted above in Figure~\ref{3scsum0615}.  It is probably a manifestation of the ``Gnevyshev gap" ({\it e.g.}, \opencite{g67}, \opencite{g77}; \opencite{s03}; \opencite{ng10}), a temporary decrease in energetic solar activity, including SEP events, often found near solar maximum.  Thus, the apparent absence of two ``$\approx$~6-month" cycles of enhanced southern SEP rate at this time may be a characteristic of the Gnevyshev gap in Cycle~24.  Finally, as in the northern hemisphere, the interval between southern SEP rate peaks appears to have shortened slightly, to $\approx$~four months, in 2014.

\begin{figure}

\begin{center}
\includegraphics*[width=10cm,angle=0]{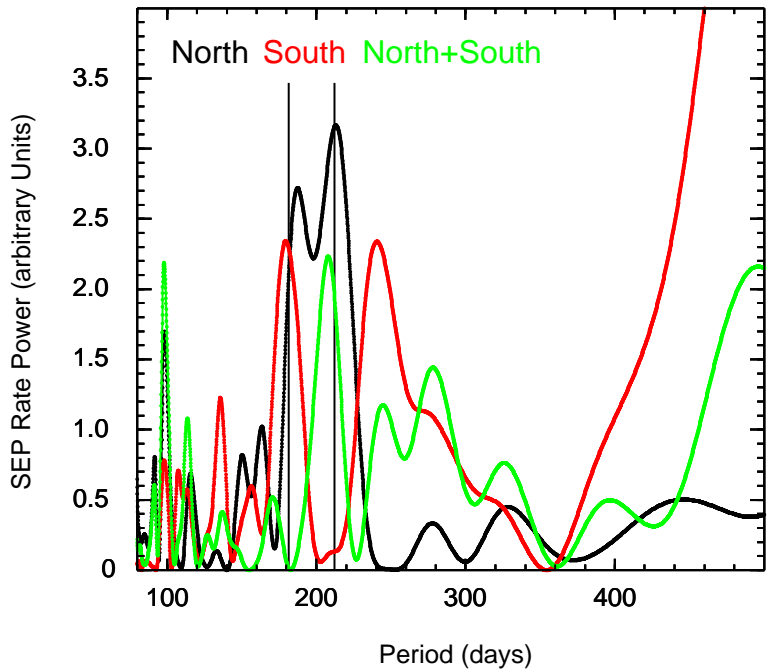}
\end{center}
\caption{Lomb-Scargle periodograms for periods of 80\,--\,500 days for the 25 MeV proton event rate in the northern hemisphere (black), southern hemisphere (red), and for the summed northern and southern rates (green) from December 2008 (start of Cycle 24) to the loss of STEREO B data.  The vertical lines indicate periods of six and seven months suggested by the observations in Figure~\ref{hemi}.}
\label{lomb}
\end{figure}

The periodicities in the SEP event rate in Figure~\ref{hemi} are certainly evident by eye, without any statistical analysis being necessary to reveal them.  Nevertheless, to demonstrate that such analysis supports this interpretation, Figure~\ref{lomb} shows Lomb-Scargle (L--S)  (\opencite{l76}; \opencite{s82}) periodograms for the northern (black), southern (red), and summed northern plus southern (green) SEP event rates from the beginning of Cycle~24 in December 2008 to the loss of STEREO B data.  The vertical black lines indicate periods of $\approx$~six and seven months suggested by the observations in Figure~\ref{hemi}.  Components close to these periods are most evident in the northern hemisphere (black), where the prominent double peak in the power spectrum has peaks at periods of 188 and 213 days.  Otherwise, the northern hemisphere shows only minor peaks (which are not statistically significant) over the period range in Figure~\ref{lomb}. The southern hemisphere spectrum (red) is more complex, and includes a large, long-period component which peaks outside the range of the figure at 506~days ($\approx17$~months) and appears to be associated with the $\approx$~18~month-variation in the event rate across the ``Gnevyshev  gap".  Otherwise, there is again a prominent component at $\approx$~six months (peak at 180~days).  Interestingly, the adjacent longer-period peak is at 240 days ($\approx$~eight months) rather than at the expected seven months. However, the two southern hemisphere intervals of seven months indicated in Figure~\ref{hemi} are both associated with enhancements in the event rate which extend over more than one rotation (for example, the rate enhancement in mid~2012 extends over $\approx$~three rotations) which may help to account for the longer period inferred from the L--S analysis.  Finally, the most prominent features in the combined rate (green) periodogram are the peaks at 208 days ($\approx$~seven months) and 114~days ($\approx$~four months).  The latter most likely arises from the out-of-phase combination of the $\approx$ 6\,--\,7 month variations in each hemisphere.  Thus, the L--S periodograms in Figure~\ref{lomb} support the conclusions inferred from Figure~\ref{hemi} that variations in the hemispheric SEP rates have periods predominantly around 6\,--7\, months.   

Comparing the top two panels of Figure~\ref{hemi}, it is clear that the SEP occurrence rates closely track variations in the sunspot numbers in the respective hemispheres, including the sunspot maximum in the northern hemisphere in 2011, the maximum in the southern hemisphere in late 2013--early 2014, and similar quasi-periodic $\approx$~six month variations in the hemispheric sunspot numbers.  Of particular note, the brief interval of enhanced southern SEP rate in mid~2012 was associated with an increase in the southern sunspot number that temporarily exceeded that in the northern hemisphere following the northern sunspot maximum and was part of the periodic variations in southern sunspot number.  Similar features are also present in the Greenwich - USAF/NOAA monthly northern and southern visible hemisphere sunspot areas (in millionths of the visible hemisphere, MH), shown in the third panel of Figure~\ref{hemi}; the sunspot areas are compiled by David Hathaway 
(\url{http://solarscience.msfc.nasa.gov/greenwch.shtml}).  The correlation between sunspot area and SEP rate will be discussed further below. 

As for the SEP rate, the $\approx$~six month variations in the hemispheric sunspot numbers and areas in Figure~\ref{hemi} can be readily identified by eye.  In addition, \inlinecite{chow15} have produced L--S periodograms of the full disk (they do not consider hemispheric) sunspot numbers and areas, and the 10.7 cm solar radio flux (a close proxy for sunspot number), for the rise of Cycle~24 from January 2009 to August 2013.  They found prominent peaks at periods of 189 and 213~days in these parameters, similar to those evident in the periodograms in Figure~\ref{lomb}.  Furthermore, their wavelet analyses of these parameters suggest that these periodicities persisted for much of their study period. 
Hence, this statistical analysis supports the picture inferred from the time series plots in Figure~\ref{hemi} of periodic variations of around six months in the hemispheric sunspot numbers and areas that persist during the rise and at least the peak phases of Cycle~24.

\begin{figure}
\begin{center}
\includegraphics*[width=7cm,angle=0]{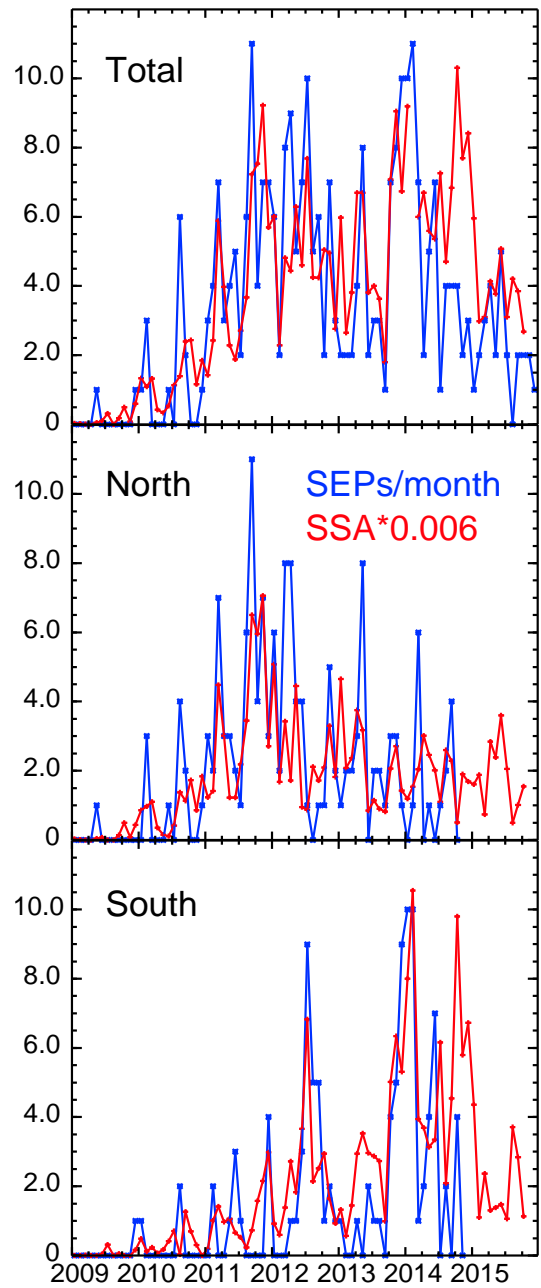}
\end{center}
\caption{Monthly rate of 25 MeV proton events observed by either STEREO spacecraft or at the Earth (blue) and monthly-averaged sunspot areas in MH scaled by a factor of 0.006 (red), for events originating on the whole Sun (top), and northern (middle) and southern (bottom) hemispheres.}
\label{scale}
\end{figure} 

The bottom panel of Figure~\ref{hemi} shows three-month running means of the hemispheric SEP occurrence rates in the top panel.  These again emphasize the close relationship between variations in the SEP rates and sunspot numbers in the same hemispheres.   The tendency, also evident in the running means, for the enhancements in SEP rate and sunspot number to be nearly in phase in each hemisphere from early in the cycle to early 2011, when they moved to be approximately in anti-phase, might be expected if the hemispheres are more tightly coupled at solar minimum than at solar maximum, as discussed by \inlinecite{ng10}.

The relationship between hemispheric SEP event rates and sunspot areas evident in Figure~\ref{hemi} is further examined in Figure~\ref{scale}.  This shows the monthly-averaged total (top panel) and northern and southern hemispheric (middle and bottom panels) sunspot areas scaled by a factor of 0.006 (red graphs) and over-plotted on the respective 25 MeV proton event rates (blue graphs).  The scaling is based on a least squares fit that indicates that the total or hemispheric number of SEP events {\it per} month is around $0.54\pm0.05$\% or $0.58\pm0.07$\%, respectively, of the total or hemispheric monthly--averaged sunspot area expressed in MH.  For 2015, the total SEP rate shown in the top panel of Figure~\ref{scale} is based only on 25~MeV proton events detected at Earth.  A number of these events evidently originated on the far side of the Sun from Earth ({\it e.g.}, they were associated with occulted type III radio bursts detected by the {\it Radio and Plasma Wave Investigation instrument} on WIND, and coronal mass ejections (CMEs) with no front side activity) but their source locations cannot be located because of limited STEREO A imaging data.  Hence, separate SEP event rates for northern and southern hemispheres in 2015 are not shown. The remarkably close relationship between variations in the total and hemispheric SSAs and SEP rates on all time scales throughout the rising and peak phases of Cycle 24, together with the hemispheric asymmetry already discussed, is very apparent in Figure~\ref{scale}.  However, there is an obvious deviation from this close relationship in late 2014, when a large, temporary increase in the southern hemisphere sunspot area was not accompanied by a corresponding increase in the southern SEP event rate.  This feature will be discussed further in the next section.

\begin{figure}

\begin{center}

\includegraphics*[width=7cm,angle=0]{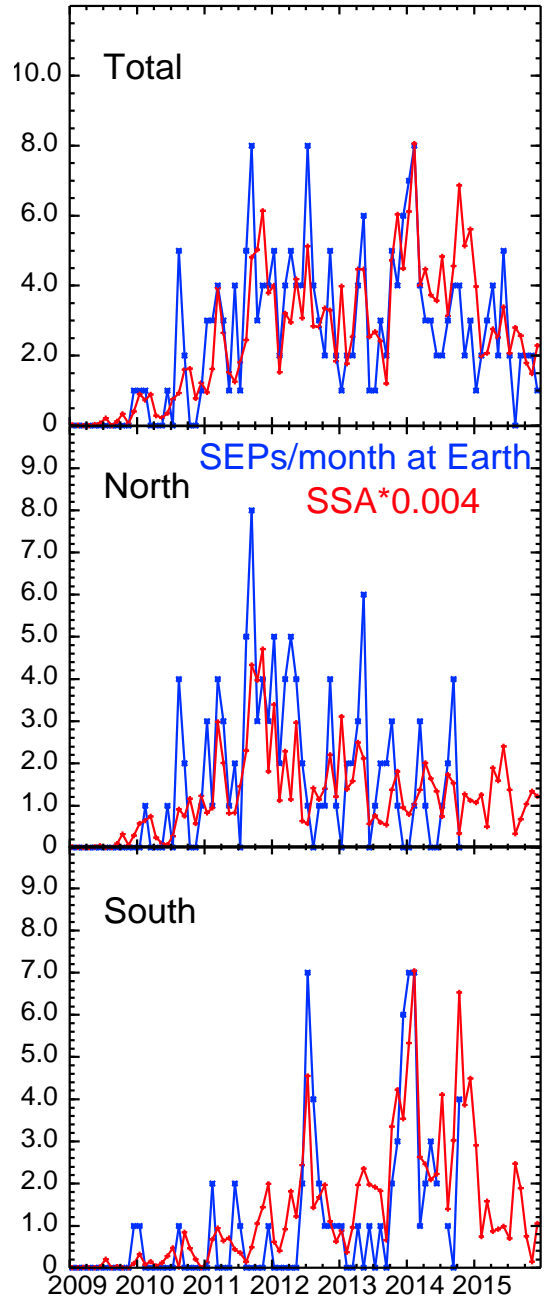}
\end{center}
\caption{Monthly rate of 25 MeV proton events observed at the Earth (blue) and monthly-averaged Sun sunspot area (MH) scaled by a factor of 0.004 (red) for the whole Sun (top) and for northern (middle) and southern (bottom) hemispheres.}
\label{scaleearth}
\end{figure}   

Figure~\ref{scaleearth} is similar to Figure~\ref{scale} (which uses SEP observations at Earth and the STEREO spacecraft) but here, the number of SEP events {\it per} month detected at Earth is shown with the total and hemispheric Sun sunspot areas scaled by a factor of 0.004 overlaid.  The scaling is based on a least squares fit that indicates that the total number of events detected at Earth is $0.37\pm0.04$\% of the total sunspot area.  Although only SEP events detected at Earth are included here (these may however also be detected at one or more STEREO spacecraft), we do use observations from STEREO to infer the hemispheric source locations of those events originating on the far side of the Sun. Hence, for the reasons discussed above, the hemispheric SEP rates are not shown for 2015. As in Figure~\ref{scale}, there is a close relationship between the total and hemispheric sunspot areas and the occurrence of 25 MeV proton events at Earth during the rising and peak phases of Cycle~24 with the exception of the interval in late 2014 when southern sunspot areas were enhanced but unexpectedly few SEP events were detected.  The difference in the sunspot area scaling in Figures~\ref{scale} and \ref{scaleearth} suggests that around a third of the proton events observed at the STEREO spacecraft are not detected at Earth, consistent with the $\approx30$\% found by \inlinecite{r14} from an event-by-event comparison.  

The close relationship between the sunspot area and SEP rate in Figures~\ref{scale} and \ref{scaleearth} suggests that estimation of the sunspot area could be the basis of forecasting the likelyhood of a 25 MeV proton event occurring, and that historical sunspot area records could help to indicate the occurrence rate of such proton events extending back to before the space era. This could be tested using observations from recent cycles where the SEP rate is known (for example, does the scaling between sunspot area and proton event rate found in Cycle 24 also hold for other cycles?), but such a test lies beyond the scope of this article.  In addition, the historical sunspot area data from different sources require careful intercalibration ({\it e.g.}, \opencite{bal09}).  Furthermore, the late 2014 period demonstrates that there are exceptions to the close relationship between sunspot area and SEP rate.

\section{Relationship of the SEP Rate with Solar and Interplanetary Parameters}

\begin{figure}

\begin{center}
\includegraphics*[width=1.0\textwidth,angle=0]{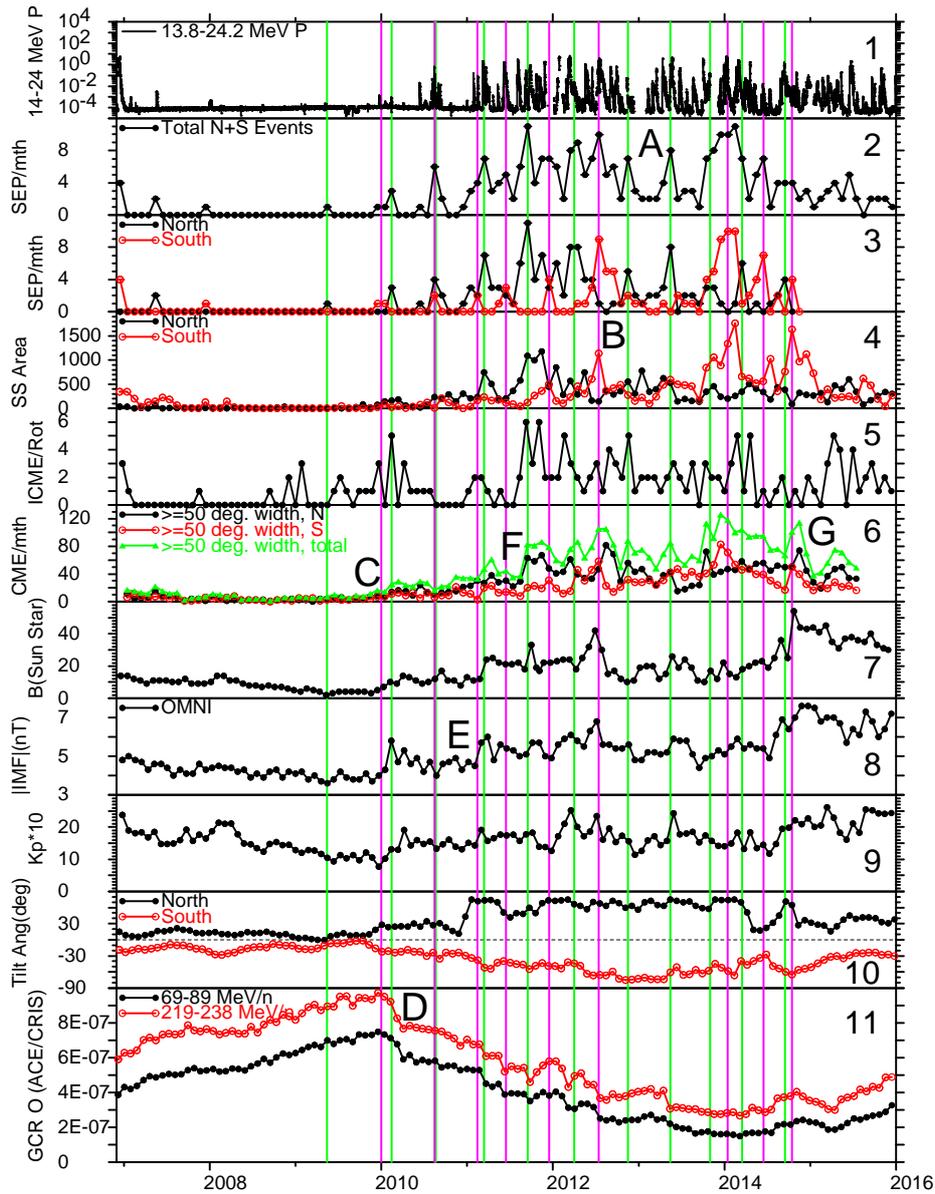}
\end{center}
\caption{Overview of various solar and interplanetary parameters from the launch of the STEREO spacecraft in October 2006 to the end of 2015. The first panel shows the 14\,--\,24~MeV solar proton intensity from the ERNE instrument on SOHO. The second and third panels show the total number of 25 MeV proton events {\it per} month based on combined SOHO and STEREO observations, and the number of these events in the northern and southern hemispheres (from Figure~\ref{hemi}), respectively.  The fourth panel shows the monthly northern and southern visible hemisphere sunspot areas, while the fifth panel shows the number of interplanetary coronal mass ejections observed at Earth {\it per} Carrington rotation.   The sixth panel shows the number of  $\ge$50$^{\circ}$-width CMEs {\it per} month, where green, black and red graphs give the total number, and the numbers of northern and southern CMEs, respectively.   The seventh panel shows Carrington rotation-averages of daily magnitudes of the Wilcox Solar Observatory mean solar magnetic field.  The eighth panel shows CR-averages of the near-Earth interplanetary magnetic field intensity, while the ninth panel shows CR-averages of the Kp*10 geomagnetic index.  The tenth panel illustrates the WSO R model north and south tilt angles of the heliospheric current sheet, while the eleventh panel shows Bartels' rotation averages of the cosmic ray oxygen intensity in two energy ranges.   The vertical green and purple lines indicate the times of peaks in the northern and southern SEP occurrence rates shown in the third panel, respectively.}
\label{solip}
\end{figure}

We now summarize, in Figure~\ref{solip}, the relationship between the occurrence of 25~MeV proton events in Cycle 24 and other solar and interplanetary parameters.  Here, the interval shown is from STEREO launch until the end of 2015, though some parameters illustrated are not available for the complete period at the time of writing.  The main conclusion is that the $\approx$~six month variations in the SEP occurrence rate are just one manifestation of the episodic development of Cycle 24 that is visible in a number of solar and interplanetary parameters.  Some other features of interest will also be noted.

 The top panel in Figure~\ref{solip} shows the 14\,--\,24~MeV solar proton intensity, in this case from the ERNE instrument on SOHO since recent observations are more complete than for the STEREO spacecraft (see Figure~\ref{3scsum0615}).  The second and third panels show the total number of 25 MeV proton events {\it per} month based on combined SOHO and STEREO observations, and the number of these events in the northern and southern hemispheres (from Figure~\ref{hemi}), respectively. As discussed above, the SEP rate in 2015 is based only on 25~MeV proton events detected at Earth, and no hemispheric rates are determined. Although the full-Sun SEP rate in the second panel does show clear temporal variations, the third panel emphasizes the additional insight obtained by separating these into their hemispheric components. Vertical green and purple lines indicate times of peaks in the northern or southern SEP event rates, respectively, as discussed in the previous section.   

The fourth panel in Figure~\ref{solip} again shows the monthly northern and southern visible hemisphere sunspot areas, while the fifth panel shows the number of interplanetary coronal mass ejections (ICMEs) observed at Earth {\it per} Carrington rotation (CR), updated from the catalog of \inlinecite{rc10} available at \url{http://www.srl.caltech.edu/ACE/ASC/DATA/level3/icmetable2.htm}.

The sixth panel in Figure~\ref{solip} shows the number of CMEs {\it per} month with plane of the sky widths $\ge$50$^{\circ}$ reported in the Coordinated Data Analysis Workshops (CDAW) CME catalog (\url{http://cdaw.gsfc.nasa.gov/CME_list/}, green graph).  The black and red graphs indicate the number of these CMEs with central position angles in the northern or southern hemispheres, respectively.  Note that halo (360$^{\circ}$ width) CMEs in the CDAW catalog are excluded here, since no position angle is provided.  However, only $\approx6$\% of the CMEs with widths $\ge$50$^{\circ}$ during the period in Figure~\ref{solip} were halo CMEs, so their omission does not significantly impact the CME rates in the figure (notwithstanding  that around a half of 25~MeV proton events are associated with halo CMEs in the CDAW catalog ({\it e.g.}, \opencite{rcs99}; \opencite{r14}; \opencite{r15}). The CME rates tend to track the sunspot area and SEP event rates in the same hemisphere, as might be expected since CMEs typically arise in association with active regions and SEPs are closely associated with CMEs.  In particular, the peaks in the SEP rate indicated by the vertical lines are frequently closely aligned with peaks in the CME rate at the Sun.  Note however that the large increase in the southern sunspot area in late 2014 is only weakly reflected in the southern CME and (as noted above) SEP event rates (and also in the ICME rate), in contrast to the similar southern sunspot area peak early in 2014.  Reasons why the southern active region 12192, the largest sunspot group in 24 years, present in October 2014, was so CME poor, including the occurrence of ``confined" flares without CMEs, have been discussed by \inlinecite{sun15} and \inlinecite{chen15}.  

The seventh panel in Figure~\ref{solip} shows Carrington rotation-averages of daily magnitudes of the Wilcox Solar Observatory (WSO) mean solar magnetic field ($B$) (\opencite{s77};  \url{http://wso. stanford.edu/meanfld/MF_timeseries.txt}), which closely track the CR-averages of the near-Earth interplanetary magnetic field (IMF) intensity (from the OMNI database; \url{http://omniweb.gsfc.nasa.gov/}) in the eighth panel; both also track the CR-averages of the Kp*10 geomagnetic index (the Kp index multiplied by 10),  from the Helmholtz Centre, Potsdam, shown in the ninth panel, reflecting the role of the magnetic field strength in modulating geomagnetic activity through the southward-directed component ({\it e.g.}, \opencite{n07} and references therein).  The tenth panel illustrates the WSO R model north and south tilt angles of the heliospheric current sheet  (\url{http://wso. stanford.edu/Tilts.html}).  Finally, the eleventh panel shows Bartels' rotation averages of the cosmic-ray oxygen-intensity in two representative energy ranges observed by the {\it Cosmic Ray Isotope Spectrometer} (CRIS) on ACE; data are from the ACE Science Center (\url{http://www.srl.caltech.edu/ACE/ASC/}).  Note that whether monthly, Carrington rotation (27.275~day) or Bartels' rotation (27.0~day) averages are used for the various parameters is not important for this discussion.)

Figure~\ref{solip} contains a wealth of observations and a full discussion is beyond the scope of this article.  However, several points of interest and relationships between various parameters, in addition to those already mentioned, are noted in the remainder of this section:

Label A in the second panel indicates the ``Gnevyshev gap" interval between the northern and southern sunspot maxima already discussed above that is characterized by temporary decreases in the rate of SEP events, sunspot area, ICME and CME rates, weakening of the mean solar field and IMF, and a reduction in Kp*10. The southern tilt angle was also at its furthest south location during the cycle, and the cosmic-ray oxygen-intensity began to recover temporarily, presumably due to the weaker solar activity leading to, for example, a weaker IMF, and fewer CMEs and ICMEs.

The 23~July 2012 solar event \cite{ru13}, which produced the most intense SEP event so far in this cycle (detected at STEREO~A; cf., Figure~\ref{3scsum0615}), occurred during the $\approx$ three-month interval in which southern activity was temporarily dominant following the northern sunspot peak (label B in the fourth panel).  The CME rate (sixth panel), mean solar field (seventh panel), and IMF intensity (eighth panel) rose during the preceding 2\,--\,3 months and peaked at around the time of the event, then declined; the peak IMF and solar field strengths were not exceeded during the rise and maximum phases of the cycle (the later period in 2014\,--\,2015 when stronger fields were present will be discussed below). Hence, this exceptional SEP event occurred at a unique time in the peak of this cycle.  Interestingly, northern-directed CMEs (black graph in the sixth panel) were more prevalent than southern-directed (red graph) around the temporary increase in southern activity in mid~2012.  The reason for this requires further investigation (for example, was there a tendency for southern CMEs to be launched non-radially or deflected northwards near the Sun at this time?). 

Turning now to the early phase of Cycle 24, the CME rate first increased around the start of 2010 (label C in the sixth panel), $\approx1$~year after smoothed sunspot minimum in December 2008.  This increase was also accompanied by enhancements in the mean solar field, IMF, and Kp*10 index from the low values characteristic of the solar minimum between Cycles 23 and 24 (sixth to ninth panels in Figure~\ref{solip}). The second of the brief periods of increased SEP occurrence during the rise of the cycle occurred around this time.  This included events from both hemispheres and, as already noted in Section~1, the first extended 25~MeV proton event observed by both STEREO spacecraft and at Earth \cite{r14}, on 22~December 2009.  The cosmic ray intensity also started its decline (label D in the eleventh panel) from the high values reached during solar minimum ({\it e.g.}, \opencite{m10}; \opencite{l13}). Typically, the onset of galactic cosmic ray (GCR) modulation in an $A<0$ (solar magnetic field inward at the North Pole) minimum is associated with an increase in the tilt angle of the heliospheric current sheet (HCS) ({\it e.g.}, \opencite{jw98}) but in this case, there was only a modest tilt angle increase just ahead of the modulation onset (tenth panel).  \inlinecite{c13} suggested instead that the modulation onset was related to diffusive processes associated with the increases in the IMF intensity and CME rate.

Around a year later (label E in the eighth panel), there were abrupt increases in the IMF intensity and mean solar field following a pole-ward jump in the northern HCS tilt angle and increase in the northern sunspot area. This was also associated with a brief interval of increased SEP occurrence (predominantly in the north but with a southern component), an increase in the CME rate, and a step down in the GCR intensity.

The CME and ICME rates both show a step up in late 2011 (label F) that was closely associated with an increase in the northern SEP rate and an increase in the northern sunspot area leading up to the first (northern) cycle maximum. There was also a step down in the GCR intensity.  

The CME rate, mean solar field, IMF intensity, and Kp (sixth to ninth panels) are all well correlated, including related steps and peaks, most associated with SEP rate peaks, throughout the period in Figure~\ref{solip} up until late 2014.  At this time, the CME rate fell but the IMF intensity rose to, and then maintained, the highest levels seen during this solar cycle (label G in the sixth panel of Figure~\ref{solip}).   A corresponding increase in the mean solar field was also observed.  This persistent increase in the IMF at a time of falling CME rates would appear to be a challenge to models ({\it e.g.}, \opencite{oc06}; \opencite{o08}) in which the increase and variations in the IMF strength during the solar cycle are associated with magnetic field lines that are dragged out into the heliosphere by ICMEs; the field lines in the legs of the ICMEs then contribute to, and become indistinguishable from, the field in the background solar wind.  A similar point is made by \inlinecite{ws15}.  They conclude that the IMF increase from mid-2014 was instead due to an increase in the strength of the equatorial dipole component of the solar magnetic field due to the emergence of large active regions with east-west dipolar moments that are in phase with the background field direction.  \inlinecite{sw15} note that similar post-maximum features were present in the previous three solar cycles.  The increase in the mean solar field at this time is also consistent with this picture \cite{sw15}, as is the detailed correlation between the IMF and mean solar field throughout Figure~\ref{solip} (and also in previous solar cycles, {\it e.g.}, \opencite{rc12}) that suggests that solar fields predominantly drive the IMF variations.

\begin{figure}
\begin{center}
\includegraphics*[width=9cm,angle=0]{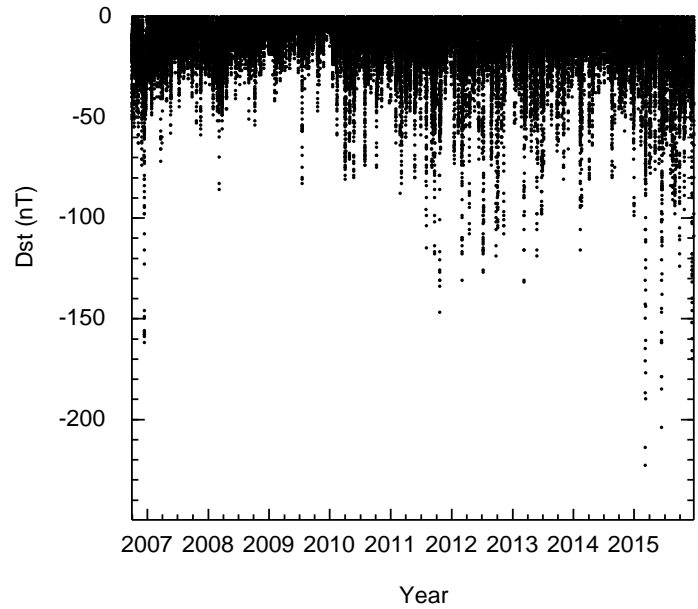}
\end{center}
\caption{Geomagnetic $Dst$ index for the period in Figure~\ref{solip}, showing the increased storm activity following the strengthening of solar and interplanetary magnetic fields from late 2014 in Figure~\ref{solip}.}
\label{dst}
\end{figure}

Geomagnetic activity also increased in association with the stronger interplanetary magnetic fields, as indicated by Kp*10 in Figure~\ref{solip}, and by the geomagnetic index $Dst$ plotted in Figure~\ref{dst} for the period in Figure~\ref{solip}.  This indicates an increase in the occurrence of enhanced geomagnetic activity in 2015, including the first two severe ($Dst\le -200$~nT) storms of Cycle 24, on 17~March and 23~June.  Both were associated with the passage of ICMEs, as is typical of such storms \cite{z07}. 

One of the most intense 25~MeV proton events so far detected in this cycle, commencing on 13~December 2014 ({\it cf.}, Figure~\ref{3scsum0615}) and reaching 53 (MeV s cm$^2$ sr)$^{-1}$ at 24.8\,--\,26.4~MeV at STEREO A, also occurred during this interval of enhanced fields.  This intensity is comparable to the initial stages of the 23~July 2012 event where, however, a further approximately order of magnitude intensity increase was localized around the passage of the related shock \cite{ru13}.  STEREO A imaged the solar activity associated with the 13~December proton event commencing at $\approx$14:05~UT in the southern active region belt (consistent with the dominant southern activity at this time in Figure~\ref{hemi}) at $\approx$ 30$^{\circ}$ to the east relative to the spacecraft, corresponding to $\approx$ 137$^{\circ}$ to the west relative to Earth, where a weaker 25 MeV proton event (peak intensity 0.01 (MeV s cm$^2$ sr)$^{-1}$) was detected. The associated CME was a full halo with a plane of the sky speed of 2222 km~s$^{-1}$ in the CDAW LASCO CME catalog.

Returning to Figure~\ref{solip}, some (but not all) of the peaks in the SEP rate and variations in other parameters are closely related with the step decreases in the GCR intensity (eleventh panel) that are typical of the ascending phase of a solar cycle ({\it e.g.}, \opencite{m98}).  The GCR steps may be related to diffusive barriers propagating away from the Sun that are driven by changes in the solar magnetic field ({\it e.g.}, \opencite{c99}; \opencite{c01}; \opencite{w02}; and references therein).  Note that the recovery in the GCR intensity that commenced in 2014 following solar maximum was temporarily interrupted late in this year, possibly associated with the strengthening interplanetary magnetic field (eighth panel) and/or the related increases in the tilt angle (tenth panel).

Did the $\approx$ six-month periodicity indicated in several parameters during the rising and peak phases of Cycle 24 extend into this period of enhanced fields following the maximum of Cycle 24?  In Figure~\ref{solip}, the ICME and CME rates exhibit a peak around April\,--\,May 2015, some 6\,--\,7 months after the prominent southern sunspot area peak in October 2014 and CME rate peak in November 2014.  There was also a small increase in the northern sunspot area around this time, as well as in the SEP rate (see also Figure~\ref{scale}).  An increase in the southern sunspot area then followed as the northern area decreased, indicating the continuing independence of the activity variations in each solar hemisphere.  So the observations suggest that such variations continued into 2015.

\section{Summary and Discussion}
As previously reported by \inlinecite{r14} using STEREO and near-Earth observations, Cycle 24 shows evidence during the early rise phase for clusters of solar energetic particle events including 25~MeV protons that occurred at intervals of $\approx6-7$ months.  By separating the SEP events according to their source hemispheres, we have demonstrated that these periodicities persisted independently in each hemisphere through the maximum of this cycle and most likely into the early declining phase.  We have also illustrated that these variations in the SEP rate are closely related to other features at the Sun and in the heliosphere, including variations in the sunspot numbers and areas in each solar hemisphere, CME and ICME rates, variations in the solar and interplanetary magnetic field strengths, and some of the downward steps in the long-term modulation of galactic cosmic rays during the rise of Cycle 24.  In particular, the close correlation between the sunspot area and the occurrence rate of 25 MeV proton events discussed in Section~2 suggests that the routine monitoring sunspot parameters provides a simple tool for forecasting the occurrence of SEP events.  If the sunspot area is small, the probability of an SEP event is low and an ``all-clear" forecast can be issued, whereas when sunspot areas are large, the probability of an event is much enhanced.  Though this is a not unexpected conclusion, the results presented here help to quantify the likely SEP rate based on the sunspot area.  However, the unusual situation in late 2014 demonstrates that occasionally, a large sunspot area may not be accompanied by the expected number of SEP events, so a ``false alarm" would result in this case.    

The observation of periodicities in solar and interplanetary phenomena is certainly not new or unique to Cycle~24.  In particular, \inlinecite{r84} recognized a $\approx$154 day periodicity in X-ray flares during Cycle~21.  Such ``near-Rieger" periodicities have also been reported by, for example: \inlinecite{l90} (who identified periods of $\approx130$--185 days in a survey of multiple sunspot cycles); \inlinecite{bc90} (in proton-producing flares); \inlinecite{g93} (geomagnetic activity);  \inlinecite{crr98} (IMF and 25 MeV proton intensity in Cycle 21, including event clustering similar to that reported here); \inlinecite{d01} (SEP events in Cycle 23); \inlinecite{h01} (anomalous cosmic ray intensity in the outer heliosphere); \inlinecite{b04} (photospheric magnetic field); \inlinecite{rc05} (SEP intensity, ICME and geomagnetic storm sudden commencement rate, hemispheric sunspot numbers, IMF); \inlinecite{l03} and \inlinecite{lara08} (coronal mass ejections); and \inlinecite{lob12} (solar type III radio bursts in Cycle 23).  As discussed by several of these papers, in particular \inlinecite{l90}, these periodicities vary in strength and period both from cycle to cycle and within a given cycle.  The observations presented in this article (and also, for example by \opencite{chow15}) indicate that Cycle 24 shows particularly clear evidence of near-Rieger periodicities in a range of solar and interplanetary phenomena during the rising and peak phases.  

The origin of the near-Rieger periodicity has been suggested to be near the surface of the Sun (\opencite{b87}; \opencite{b88}; \opencite{l00}), due to changes in the rate of
solar magnetic flux emergence ({\it e.g.}, \opencite{crr98}; \opencite{o98}; \opencite{b02}) or a ``global" phenomenon ({\it e.g.}, \opencite{bs87}; \opencite{w92}), while \inlinecite{l90} noted an association with complex active regions containing large sunspots (``super-active regions").  
Recently, \inlinecite{m15} have suggested that the variability in the number of flares, CMEs, particle events, and other solar and interplanetary phenomena is driven by surges of magnetism from activity bands of the 22-year solar cycle which are in turn driven by the deep solar interior.  They characterize this variability as quasi-annual, and suggest that the Rieger-like periods are what they term ``hybrid" periodicities that arise from quasi-annual periodicities that are out of phase in each solar hemisphere such that, depending on the phasing, the Rieger periods are more or less visible in a whole-Sun time series.  However, our observations point to a different picture in which $\approx$~six-month near-Rieger periodicities exist and evolve in each hemisphere independently (though the tendency for the variations to be either in phase or anti-phase suggests that there may be some coupling), at least during the interval of Cycle 24 discussed here.  We do not see evidence of prominent, quasi-annual periodicities in each hemisphere ({\it c.f.}, the lack of an annual ($\approx365$~day period) signal in the L--S periodograms in Figure~\ref{lomb}) that then combine to produce the shorter-term Rieger periodicities.  Thus, the observations discussed here do not appear to support the interpretation of the Rieger periodicities proposed by \inlinecite{m15}.  On the other hand, the combination of hemispheric periodicities to give shorter-period full-Sun variations in the SEP rate is suggested by the full-Sun L--S periodogram in Figure~\ref{lomb} and by the north plus south SEP rate in the second panel of Figure~\ref{solip}.  This suggests that a reconciliation with the \inlinecite{m15} picture may be possible if the fundamental hemispheric variations during the ascending and maximum phases of Cycle 24 are not quasi-annual but have periods of 6\,--\,7 months, and these then combine to give shorter period variations, in this case with periods of $\approx100$~days, below the previously-reported range of the Rieger-like periods \cite{l90}. 

Although periodicities in solar and interplanetary phenomena lying between the solar rotation period and $\approx$~1~year have been identified for over three decades, they have not been incorporated into understanding and predicting space weather on time scales of around a few months to a year.  One factor may be that such periodicities are often regarded as mere statistical curiosities (not withstanding the clear periodicities exhibited by the Sun on other time scales), lacking a clear physical explanation.  In addition, they tend to be ephemeral, being most obvious only during certain intervals of particular cycles, and with periods that may change with time.  Maybe the particularly clear periodicities observed during Cycle 24 as discussed here, together with a convincing link with phenomena deep in the Sun, such as suggested by \inlinecite{m15}, will invigorate interest in this topic and lead to an improved understanding of these variations and their consequences throughout the heliosphere.      
   
%
\begin{acks}
 We thank the many researchers who have compiled the various data sets used in this article.  The STEREO {\it High Energy Telescope} data are available at \url{http://www.srl.caltech.edu/STEREO/ Public/HET_public.html}.  The ERNE data are from the Space Research Laboratory at the University of Turku (\url{http://www.srl.utu.fi/erne_data/}).  This work was supported by the NASA Living With a Star Science program as part of the activities of the Focused Science Team ``Physics-based methods to predict connectivity of SEP sources to points in the inner heliosphere, tested by location, timing, and longitudinal separation of SEPs".
\end{acks}

{\bf Disclosure of Potential Conflicts of Interest}
The authors declare that they have no conflicts of interest.


\end{article} 
\end{document}